\documentclass[conference,a4paper]{IEEEtran}
\IEEEoverridecommandlockouts
\usepackage{cite}
\usepackage{amsmath,amssymb,amsfonts}

\usepackage{mathtools}

\usepackage{cuted}
\usepackage{graphicx}
\usepackage{textcomp}
\usepackage{xcolor}

\usepackage{amsmath, amssymb, amsthm,algorithm}
\usepackage[latin1]{inputenc}

\usepackage{latexsym}
\usepackage{graphics,epsfig}
\usepackage{amsfonts}
\usepackage{booktabs}
\usepackage{color}
\usepackage{multirow}
\usepackage[justification=centering]{caption}
\usepackage{dsfont}
\usepackage{graphicx}
\usepackage{adjustbox}
\usepackage{kantlipsum}
\usepackage[caption=false,font=footnotesize]{subfig}
\usepackage{cases}
\usepackage{url}
\usepackage{tabu}
\usepackage{makecell}
\usepackage{soul}

\usepackage{array}
\usepackage{algpseudocode}
\usepackage{mathtools,lipsum,cuted}

\usepackage[justification=centering]{caption}

\usepackage[caption=false,font=footnotesize]{subfig}
\usepackage[T1]{fontenc}
\usepackage{mwe}
\usepackage{subfig}

\newlength{\tempheight}
\newlength{\tempwidth}

\newcommand{\rowname}[1]% #1 = text
{\rotatebox{90}{\makebox[\tempheight][c]{#1}}}

\newcommand{\columnname}[1]% #1 = text
{\makebox[\tempwidth][c]{#1}}

\makeatletter
\def\BState{\State\hskip-\ALG@thistlm}
\makeatother

\newcounter{example}[section]

\usepackage[english]{babel}
\usepackage{verbatim}

\usepackage[english]{babel}

\theoremstyle{plain}

\theoremstyle{remark}

\def\BibTeX{{\rm B\kern-.05em{\sc i\kern-.025em b}\kern-.08em
    T\kern-.1667em\lower.7ex\hbox{E}\kern-.125emX}}
\begin{document}

\title{Scaling Serverless Functions in Edge Networks: \\A Reinforcement Learning Approach}

\author{\IEEEauthorblockN{Mounir Bensalem$^{*}$,  Erkan Ipek$^{*}$ and Admela Jukan$^{*}$}
\IEEEauthorblockA{$^{*}$Technische Universit\"at Braunschweig, Germany;
\{mounir.bensalem, e.ipek, a.jukan\}@tu-bs.de}
%$^{+}$CARIAD SE, Germany;
%anna.engelmann@cariad.technology

}

\maketitle

\begin{abstract}

With rapid advances in containerization techniques, the serverless computing model is becoming a valid candidate execution model in edge networking, similar to the widely used cloud model for applications that are stateless, single purpose and event-driven, and in particular for delay-sensitive applications. One of the cloud serverless processes, i.e., the auto-scaling mechanism, cannot be however directly applied at the edge, due to the distributed nature of edge nodes, the difficulty of optimal resource allocation, and the delay sensitivity of workloads. We propose a solution to the auto-scaling problem by applying reinforcement learning (RL) approach to solving problem of efficient scaling and resource  allocation of serverless functions in edge networks. We compare RL and Deep RL algorithms with empirical, monitoring-based heuristics, considering delay-sensitive applications. The simulation results shows that RL algorithm outperforms the standard, monitoring-based algorithms in terms of total delay of function requests, while achieving an improvement in delay performance by up to 50\%. 

\end{abstract}


\section{Introduction}
%%%%%%%%%%%%%%%%%%%%%%%%%%%%%%%%%%%%%%%%%%%%%%
In serverless computing, function scaling is a key process to manage resource allocation by creating and removing function instances/replicas when functions are requested or idle for a certain duration. In the cloud, function scaling is managed in a centralized fashion as implemented in a few well known commercial serverless platforms, such as Amazon AWS Lambda or Google Cloud Functions, as well as in the related open source tools, like OpenFaaS or Apache OpenWhisk. As recognized by a few recent studies,  the serverless platforms can be applied also in the context of edge networks, such as in works \cite{palade2019evaluation, baresi2019towards, Carpio2022, li2022kneescale}. Furthermore, recent work investigated problems related to execution model  \cite{hall2019execution},  resource provisioning \cite{ascigil2021resource}, placement of resources \cite{Bensalem2020}, and resource scaling \cite{li2022kneescale, zhang2022adaptive, bensalem2023towards}. The state of the art work adopts scaling methods based on monitoring of the function arrivals, while the resource allocation uses periodic collection of telemetry data and various optimizations. 

What is currently missing in the edge networking context are the novel auto-scaling mechanisms, due to the distributed nature of edge nodes, the difficulty of optimal resource allocation, and the delay sensitivity of workloads typical for the edge context. The current function scaling methods are based on monitoring of the function arrivals only, and do not consider the network states, which is critical. On the other hand, resource allocation methods that depend on periodic collection of telemetry and optimal linear programming models are not practical for real time scenarios. In our previous work \cite{bensalem2023towards}, we studied the optimality of serverless function scaling using Semi-Markov Decision Process-based (SMDP) theoretical models, but also these are rather computationally demanding, and do not consider delay-sensitive applications. Delay constrained functions and applications, such as image recognition and anomaly detection,  require a guaranteed response time. Since most serverless platforms were developed for the cloud, delay sensitive workloads are out of scope, and hence not application in the edge context. 

%%%%%%%%%%%%%%%%%%%%%%%%%%%%%%
We propose a novel and practical solution to the auto-scaling problem by applying reinforcement learning (RL) and its extension to deep RL in edge networks. We compare our  algorithms with empirical , monitoring-based heuristics, while considering delay-sensitive applications. It should be noted that recent work  \cite{zhang2022adaptive} used RL algorithm to characterize the service profile when making the scaling decision in the cloud. Furthermore,  \cite{yao2023performance} uses deep reinforcement learning (DRL),  to achieving distributed function allocation at the edge. Our focus is the consideration of delay constraints, which is novel. In our approach, we prioritize the delay constraints after computing all the available allocation possibilities, thus reaching the best possible scaling solution. Furthermore, we consider capacity constraints and queuing and transmission delays. We also uniquely consider concurrency of service requests over resources.  We compare simulation results to standard, monitoring-based algorithms in terms of total delay of function requests.  RL outperforms monitoring-based methods used by default in today's serverless platforms by decreasing the average delay of function requests, and enhancing the satisfaction rate of delay constraints for all request loads, and up to 50\% for certain request loads. Our method is practically relevant as it can be easily implemented in open source serverless platforms to scale functions, as it only need to collect data about the delay of each function request, which the container orchestrator can easily provide.

%%%%%%%%%%%%%%%%%%%%%%%%%%%%%%%%%%%%%%%%%%%%%%%%%%%%%%%

The rest of the paper is organized as follows. 
 Section \ref{sec:sysmodel} introduces the system model. Section \ref{sec:rl} presents the RL and DRL auto-scaling approaches, and 
Section \ref{sec:results} evaluates the performance. Section
\ref{sec:conclusion} concludes the paper.

\section{System Model}\label{sec:sysmodel}%\hl{Done}

\begin{figure}[!t]
  \centering
  \includegraphics[width=0.92\columnwidth]{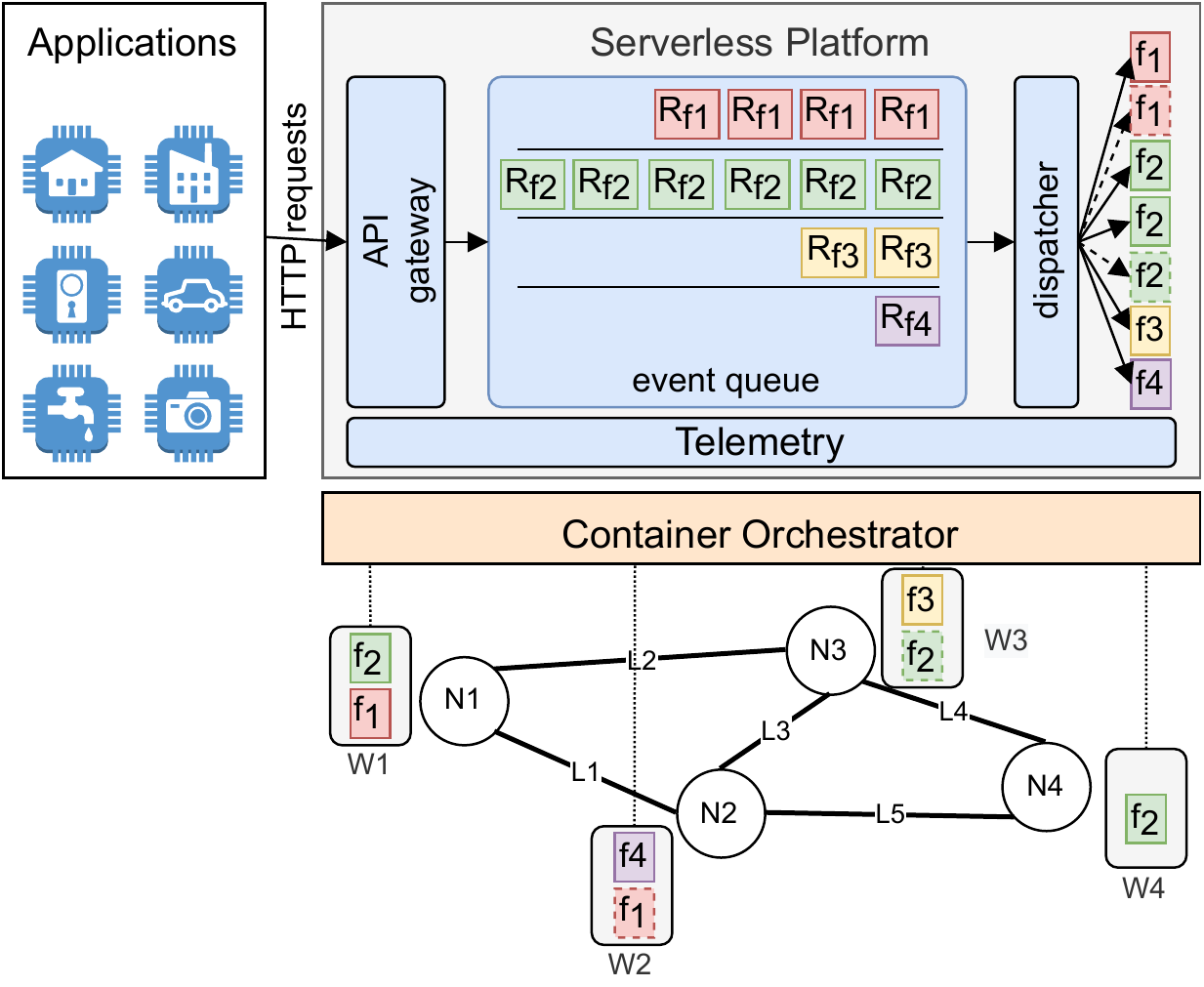}
  \caption{Serverless system at the edge}
  \label{fig:system}
\end{figure}

Fig. \ref{fig:system} illustrates the system model which includes an example physical infrastructure, here consisting of 4 edge computing nodes (W1, W2, W3 and W4), a container orchestrator, generally Kubernetes, a serverless platform, and several applications. The serverless platform (such as OpenFaaS) typically consists of an application programming interface (API) gateway, event queue, dispatcher and telemetry modules; the latter has access to all resources in the network. Applications/Users can request functions via HTTP requests through the API gateway. Functions can be of different types, depending on the application, and for each function type the serverless platform creates an event queue where the said function need to wait (queue) to be processed.

The dispatcher is the module that controls the scaling mechanism, which is the main focus of our paper. The implementation of the dispatcher typically considers the type of function, number of requests per function, and the resource utilization policy. The default dispatcher scheme used in openFaaS checks periodically the arrival rate  (or load) of each function; as a consequence, it  scales-up the number of replicas whenever it exceeds a certain threshold, or scales-down the number of replicas whenever the queue is emptied. The telemetry component collects information about the arrival event and the queue status. After making the scaling decision by the serverless platform, this decision is handled by the container orchestrator to place functions accordingly in the edge computing network. Joint scheme of scaling and placement can also be used to suggest an allocation decision to the container orchestrator, which allows us to propose a combined approach in our paper. Our model additionally assumes that functions can be requested with a specific delay constraint, such that delay sensitive applications can be considered. 

 \subsection{Problem Formulation}
%%%%%%%%%%%%%%%%%%%%%%%%%%%%%%%%%%%%%%%%%%%%%%%%%%%%%
We consider a single master multi-worker deployment, including a set of edge nodes  $\mathcal{E}=\{E_{1},..,E_{n},...,E_{N}\}$, where $E_{n}$ represents the $n^{\text{th}}$ edge node. We assume that each edge node is constrained by a certain amount of capacity modeled as a number of CPU units $C_{n}$.  
We consider a set of function classes denoted as $\mathcal{K}=\{1,...k,...,K\}$, where each function with class $k$ requires the same amount of resources $b_k$ CPU units.  Let $\Xi=\{ \xi_1,..., \xi_k,..., \xi_K\}$ be a set of function delay constraints, where each function request $f_k$ of a function class $k$ is assumed to have a certain delay requirement $\xi_k$. 
%%%%%%%%%%%%%%%%%%%%%%%%%%%%%%%%%%%%%%%%%%%%%%%%%%%%%
 The arrival and service  processes of function requests of
class $k$ are modeled  as a Poisson process with rates $\lambda_k$, and $\mu_k$, respectively. We assume that all function requests are queued in a buffer with infinite capacity. 
\subsection{Allocation model}
Each function of class $k$ requires an amount of resource that can be served by at least one available edge node, such that:
\begin{equation}\label{cpuConst}
 b_k\leq max_{\forall n: n\in[1,N]}\{C_n\},  \forall k\in [1,K]
 \end{equation}
We denote by $\delta_{k}(n)$ the number of replicas from function of class $k$ allocated in edge node $E_n$. The resource allocation has the capacity constraints, which can be given as:
\begin{equation}\label{CapConst}
 \sum_{k=1}^{K}b_k \delta_{k}(n) \leq C_n, \forall n\in [1,N]
\end{equation}
\subsection{Delay Model}
Unique to our work is the delay model, whereby we considers three types of delays:  processing delay, transmission delay,  and queuing delay.  \emph{The processing delay}  $D_p^{k, n}$ of a function $f(k)$ of class $k$ deployed through a pod in node $E_n$ is constant. This is due to the fact that serverless functions are commonly used for single purpose processing, where the input and output data are always the same in terms of size and the processing operations are the same. \emph{Transmission and propagation times}, can also be defined as a constant parameter since they depend on the distance between worker (edge) nodes and the so-called master node (e.g., in Kubernetes).  In a single master multi-worker deployment, the routing and path computation is assumed to be managed by the container orchestrator (e.g. Kubernetes). We denote by $D_t^{k, n}$ the transmission delay of a function request $f(k)$ between master node and worker node $E_n$. Let us hence focus on the total total queuing delay.

\emph{The total queuing delay} $D_q^{f(k), k}$ of a function $f(k)$ of class $k$ is restricted to the delay caused by the scaler when it decides to put a function request in the queue. The queuing delay is measured as the difference between the arrival request time plus the transmission and the processing time and the departure request time. Finally, the total delay $D^k$for each function request $f(k)$ is given as:
\begin{equation}\label{totdelay}
 D^{f(k), k}= D_p^k + D_t^{k} + D_q^{f(k), k}, \forall k\in [1,K]
\end{equation}

Time sensitive applications impose delay constraints, i.e., 
\begin{equation}\label{delayConst}
 D^{f(k), k}\leq \xi_k, \forall k\in [1,K]
\end{equation}

\section{RL and DNN-based Scaling Model}\label{sec:rl}%\hl{Done}
In this section, we first propose a basic RL-based  algorithm to provide the best possible scaling decision of serverless functions,  with actions of scaling the functions up or down by creating and removing replicas considering  the processing and queueing costs while satisfying delay requirements. After that, we extend this model using Deep Reinforcement Learning.

\subsection{Reinforcement Learning Scaling Model}
The basic RL model is defined by  an agent (the scaler), a state space $S$, an action space $A$, a system reward $R$, and an environment.   The agent (scaler) selects an action that changes the resource allocation of functions through the function provider, which results a change of the network state into a new state. The RL agent evaluates the total delays of function requests, i.e. round trip time, and assigns a reward to the decision. RL model uses a Q-table to store at each step the agent's needed knowledge for the decision making process, including the state, action, reward and the next state.

%%%%%%%%%%%%%%%%%%%%%%%%%%%%%%%%%%%%%%%%%%%%%%%%%%%%%
The system state $s$ at time $t$ for the RL model is represented by the node allocation decision, the queue length, the event type that can happen in the system, and the type of the requested function: 
\begin{equation}\label{eq:state}
\begin{split}
S=\{s|s=& (\Delta, Q, e, f_k)\},
\end{split}
\end{equation}
 where  a set $\Delta=\delta_{1},...,\delta_{k},...,\delta_{K}$ encodes the nodes availability for each class of functions, where the variable $\delta_{k}$ is a binary variable equal to $1$ if there is enough capacity to replicate a function of class $k$ in the edge computing system and $0$ otherwise. $Q=\{ Q_1,...,Q_K\}$  denotes the function request queue length vector. A binary variable $e$  indicating the type of the event that occurs in the system, is set to $0$ for arrival events and $1$ for departure events. A function identification $f_k$ is an integer variable that defines the class $k$ of a function.
%%%%%%%%%%%%%%%%%%%%%%%%%%%%%%%%%%%%%%%%%%%%%%%%%%%%%

The RL agent has a set of actions $a(s)$ to take at every new event in the system (arrival or departure): to allocate a function replicas in a specific node, to place a function request in the queue, or to remove a function replicas. In our RL model, the action variable is defined as:
\begin{equation}\label{eq:action}
A(s)= \begin{cases}
      \{0, 1,...,n,..., N \},\;&  e\in Ar \\
     \{-1, 0 \},\; & e\in D \\
    \end{cases}
\end{equation}

where $a(s)=n,\forall k\in \{1,...,K\}$ when a function  of class $k$ is replicated in edge node $E_n$, $a(s)=-1,\forall k\in \{1,...,K\}$ when a function of class $k$ is removed from the system, $a(s)=0$ denotes the action of queuing a function request of any class $k$ for function arrival events, and the queue update for  function request departure. 

%%%%%%%%%%%%%%%%%%%%%%%%%%%%%%%%%%%%%%%%%%%%%%%%%%%%%
 
At every system state $s$ and after taking an action $a(s)$, the scaling agent obtains a reward in order to gain  knowledge and adapt its decision making process accordingly. The reward function $R(s, a)$  used in our RL model considers the transmission delay value and the delay constraint to encourage the agent to select actions that satisfy delay constraints and achieving a lower average delay while maximizing the time satisfaction rate. 
The reward is defined as following:

\begin{equation}
	R(s, a) = 
	\left\{
	\begin{array}{ll}
		r_1, & \mbox{if } \xi_k \mbox{ is satisfied and } \psi = 0, \\
		r_1 \cdot \frac{w_1}{\psi}, & \mbox{if } \xi_k \mbox{ is satisfied and } \psi \neq 0, \\
		r_2, & \mbox{if } \xi_k \mbox{ is not satisfied and } \psi = 0, \\
		r_2 \cdot \frac{w_2}{\psi}, & \mbox{if } \xi_k \mbox{ is not satisfied and } \psi \neq 0.
	\end{array}
	\right. 
	\label{eq:rewardfunction}
\end{equation}
where $a$ is action, $\xi_k$ is function delay constraints, $\psi$ is transmission delay between master node and worker node allocated through action $a$, and $wr_1$ and $wr_2$ are weights. 

%%%%%%%%%%%%%%%%%%%%%%%%%%%%%%%%%%%%%%%%%%%%%%%%%%%%%

The detailed reinforcement learning process is given by \textbf{Algorithm \ref{alg:QLalgorithm}}.  The goal of the algorithm is to select an action, as defined in eq. (\ref{eq:action}). In the case of warm-up, where previous actions were taken or the RL model still did not learn yet how to accuratly choose actions, an exploration phase is needed. We consider an $\epsilon$-greedy approach to explore the search space and try new actions. We denote by $\epsilon$ the exploration probability, where at each step the RL agent  chooses  randomly an action with probability   $\epsilon$ and uses the accumulated knowledge with probability $1-\epsilon$. The learning process is repeated several times in form of episodes where the exploration-exploitation factor is decaying linearly. At every episode, the agent checks the state of the environment defined in eq. (\ref{eq:state}) to take an action $a$ using the \textit{SelectAvailableAction($s$)} function, which chooses an available action with the highest q-value, considering the $\epsilon$-greedy approach.  After that, The RL agent updates the Q-table using the following equation:

\begin{equation}\label{eq:q_learning}
\mathcal{Q}_t(s, a) = (1 - \alpha) \cdot \mathcal{Q}_t(s, a) + \alpha \cdot (R(s, a) + \gamma \cdot \max_{a'} \mathcal{Q}_t(s', a'))
\end{equation}
where $\alpha$ and $\gamma$ denote the learning rate and discount factor, respectively. We set $\alpha = 0.01$, $\gamma = 0.95$, the initial exploration probability $\epsilon=1$, and the decay$=0.98$.

\begin{algorithm}
\footnotesize %\small
	\caption{RL based scaling algorithm}
	\label{alg:QLalgorithm}
	\begin{algorithmic}[1]
		\State \textbf{Input: } events, $b_k$,  $\mu_k$,  $\xi_k$, network state ($\Delta$, $Q$, $C$)
		\State \textbf{Initialization: }$\mathcal{Q}_t$, $\epsilon$, $\alpha$, $\gamma$, episodes, decay
		\For {each episode }{
		
		\If{episode $> 0.1 \cdot$ episodes}
		    
			$\epsilon$ $\leftarrow$ decay $\cdot \epsilon$ \Comment{update exploration-exploitation rates}
		\EndIf
			\For {each event}
			{
				
				$s$ $\leftarrow$ ($\Delta$, $Q$, $e$, $f$), \;\;\;   $a$ = SelectAvailableAction($s$)

				$\Delta'$, $Q'$, $e'$, $f'$ $\leftarrow$ Find next state parameters
				
				$s'$ $\leftarrow$ ($\Delta'$, $Q'$, $e'$, $f'$)

				$\mathcal{Q}_t$ $\leftarrow$ UpdateQTable($s$, $a$, $s'$)
				
				$s$ $\leftarrow$ $s'$
		
			}
			\EndFor
		}
		\EndFor
		
		\Function{SelectAvailableAction}{$s$}
			\State actions = CheckAvailableActions($s$)
			\If{event is an arrival}
				\If{$s$ in $\mathcal{Q}_t$ }  \If{random $<$ $\epsilon$}   \Comment{$\epsilon$-greedy approach}
					\State Select randomly from actions
					\Else
						\State Select action with highest Q value from $\mathcal{Q}_t$ that exists in actions
					\EndIf
				\Else
					\State Select  randomly from  actions
				\EndIf
			\EndIf
		\EndFunction
		
		\Function{UpdateQTable}{$s$, $a$, $s'$}
			\If{$s$ not in $\mathcal{Q}_t$}
				\State   $\mathcal{Q}_t(s, a)\leftarrow 0$
			\EndIf
			%\State Find the maximum Q-value for the next state $s'$
			%\State Update the Q-value for the current state $s$ and action $a$ 
			
			$r$ $\leftarrow$ Calculate reward using (\ref{eq:rewardfunction}) \;\;\;  
				
			\State $\mathcal{Q}_t(s, a) \leftarrow (1 - \alpha) \cdot \mathcal{Q}_t(s, a) + \alpha \cdot (r + \gamma \cdot \max_{a'} \mathcal{Q}_t(s', a'))$
		\EndFunction	
		
	\end{algorithmic}
\end{algorithm}

\subsection{Deep Reinforcement Learning Scaling Model}\label{sec:drl}%\hl{in progress}
\begin{figure}[!t]
  \centering
  \includegraphics[width=0.5\columnwidth]{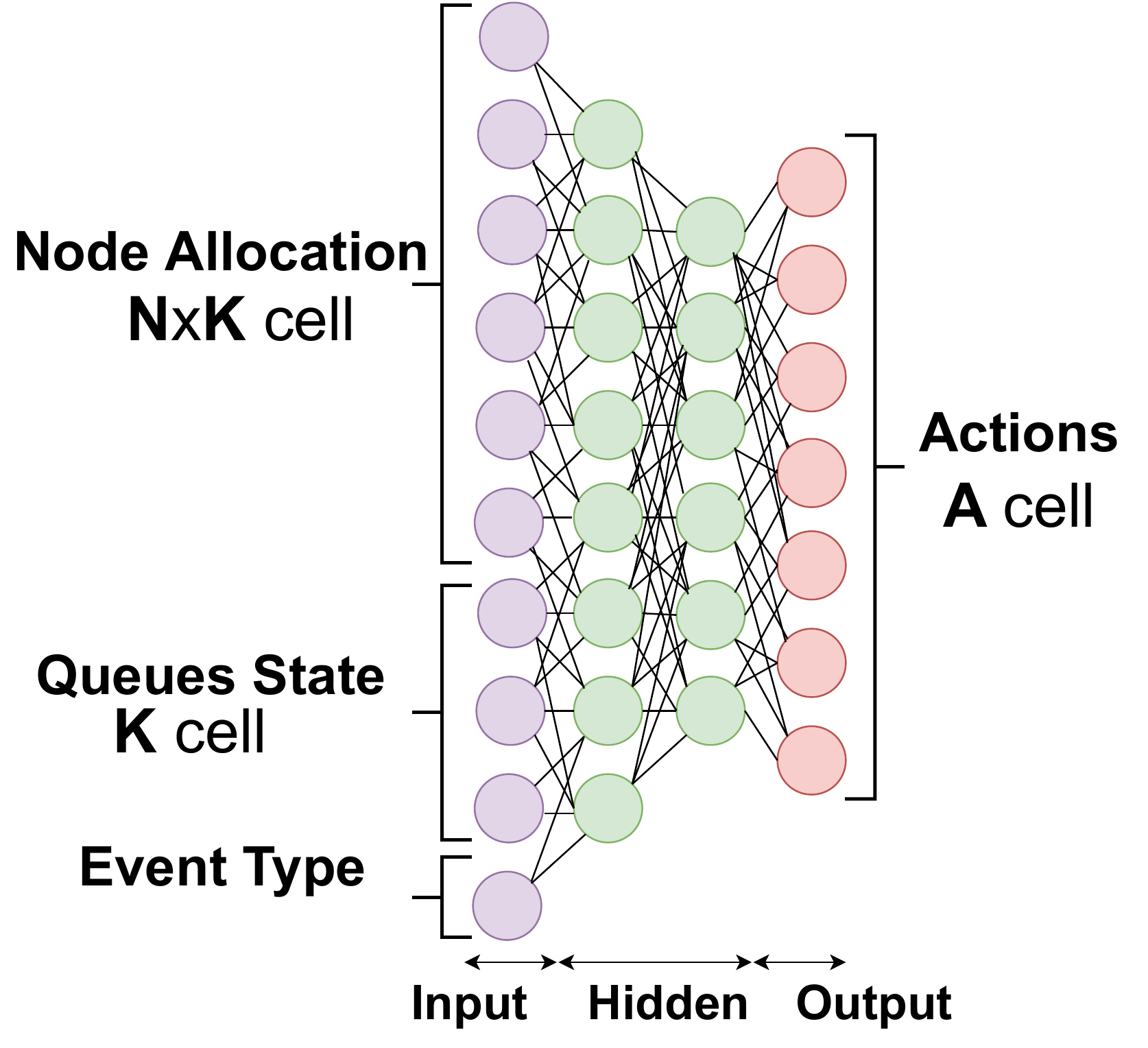}
  \caption{DNN architecture for mapping network state to actions}
  \label{fig:nn}
\end{figure}
%\vspace{-0.2cm}
We now propose an extension to the basic Q-learning scheme previously proposed, using Deep Reinforcement Learning, which employ Deep Neural Network (DNN) to estimate the $q$-values, i.e.  Q-table $\mathcal{Q}_t$. After investigation of various possible input representation, we adopt the following state definition:
\begin{equation}\label{eq:DRLstate}
\begin{split}
s=\{s|s=& (\Delta^{*}, Q, e)\},
\end{split}
\end{equation}
 where   $\Delta^{*}$ represents an $\mathbb{R}^2$ vector encoding the  number of replicas of each function of class $k$ in each node $E_n$,  $Q$ and $e$ have the same definition from eq. (\ref{eq:state}). \\
The state definition is used to define the input of DNN architecture illustrated in Fig. \ref{fig:nn}, while the output is defined by the actions defined in eq. (\ref{eq:action}).

Similar to RL algorithm, DRL uses $\epsilon$-greedy approach to explore random actions with probability $\epsilon$, where the exploration factor is decaying from an episode to another. The DNN model is initially created based on the size of the network and the number of functions to be used to predict the best scaling and placement action. At every step, the DNN model is used to estimate q-values of every possible action instead of using the Q-learning equation eq. (\ref{eq:q_learning}).  The knowledge learnt from previous experiences is stored in a replay memory $\mathcal{M}$ with a size $|\mathcal{M}|$, and update the DNN parameters using a mini-batch $\mathcal{B}$ of experiences from $\mathcal{M}$ every $U$ events. %\hl{talk about resource allocation}
\begin{algorithm}
\footnotesize %\small
	\caption{DRL based scaling algorithm}
	\label{alg:DQLalgorithm}
	\begin{algorithmic}[1]
		\State \textbf{Input: } events,  $b_k$,   $\mu_k$,  $\xi_k$, network state ($\Delta^{*}$, $Q$, $C$)
		\State \textbf{Initialization: }$\epsilon$, $\alpha$, $\gamma$, episodes, replay memory $\mathcal{M}$, batch size $\mathcal{B}$,  model update frequency $U$
		
		 \State model $\leftarrow$ Create neural network  model using $|\Delta^{*}|$ and $|Q|$ 
		\For {each episode }{
		\If{episode $> 0.1 \cdot$ episodes}
		    
			$\epsilon$ $\leftarrow$ decay $\cdot \epsilon$ \Comment{update exploration-exploitation rates}
		\EndIf
			\For {each event}
			{
				
				$s$ $\leftarrow$ ($\Delta^{*}$, $Q$, $e$), \;\;\;   
				
				$\mathcal{Q}_t(s)$ $\leftarrow$  model($s$)  \Comment{predict all q-values for state $s$ using the model}
				
				$a$ = SelectAvailableAction($s$, $\mathcal{Q}_t(s)$)

				$\Delta^{*'}$, $Q'$, $e'$, $\leftarrow$ Find next state parameters
				
				$s'$ $\leftarrow$ ($\Delta^{*'}$, $Q'$, $e'$, $f'$)

				model $\leftarrow$ UpdateDRL($s$, $a$, $s'$)
				
				$s$ $\leftarrow$ $s'$
		
			}
			\EndFor
		}
		\EndFor
		
		\Function{SelectAvailableAction}{$s$, $\mathcal{Q}_t(s)$}
			\State actions = CheckAvailableActions($s$)
		\If{event is an arrival} \If{random $<$ $\epsilon$}    \Comment{$\epsilon$-greedy approach}
					\State Select randomly from  actions
				\Else
					\State Select  action with highest q-value from $\mathcal{Q}_t(s)$  that exists in actions
				\EndIf
			\Else
				\State Select randomly from  actions
			\EndIf
		
		\EndFunction
		
		\Function{UpdateDRL}{$s$, $a$, $s'$}
			\State Add experience ($s$, $a$, $r$, $s'$)   to the replay memory  $\mathcal{M}$ 
			\If{ every  $U$ event }
				\State data  $\leftarrow$ Get random $\mathcal{B}$ experience from $\mathcal{M}$
				\State Train  model with sampled  data
			\EndIf

		\EndFunction	
	\end{algorithmic}
\end{algorithm}

\section{Performance Evaluation}\label{sec:results}%\hl{in progress}
We study the performance of our RL and DRL methods through simulations. We evaluate the best settings leading to convergence, and compare ours to the algorithms used in an open source serverless platform by default, e.g., in  OpenFaaS. 
\subsection{Simulation Setup}
We used an event based simulator running up to 100000 events with severals seeds  to simulate the network behavior and  validate our results. We evaluate a network that consists of $N=10$ edge nodes. The transmission time between master node and worker nodes is generated using a uniform distribution of interval $[0,30]$. We assume that users can request $K=5$ types of functions. 
 We generate all %function 
 arrival requests using exponential distribution with a mean value $\lambda_k=2.5-4$. The service time of each function request follows an exponential distribution with a mean values $\mu_k=5, 6, 7.5, 10, 13$. We assume the function requirements $b_k=k $ cpu units and  delay constraints $\xi_k= 20, 23, 26, 29, 32 $ (ms).

\subsection{RL and DRL Settings}
We first evaluate the learning performance of RL and DRL algorithms in order to choose the best hyperparameter setups. 
\subsubsection{RL} We set $\alpha$= 0.01, $\gamma$ = 0.95, nb. of episodes $=100$, Initial $\epsilon=1$, and Decay $=0.98$.
\subsubsection{DRL} Number of episodes $=10$,  $\mathcal{B}=1280$, $U=2500$, size of DNN hidden layer 1 $=32$,   hidden layer 2 $=16$.

\begin{figure*}
\centering
\includegraphics[width=0.32\textwidth]{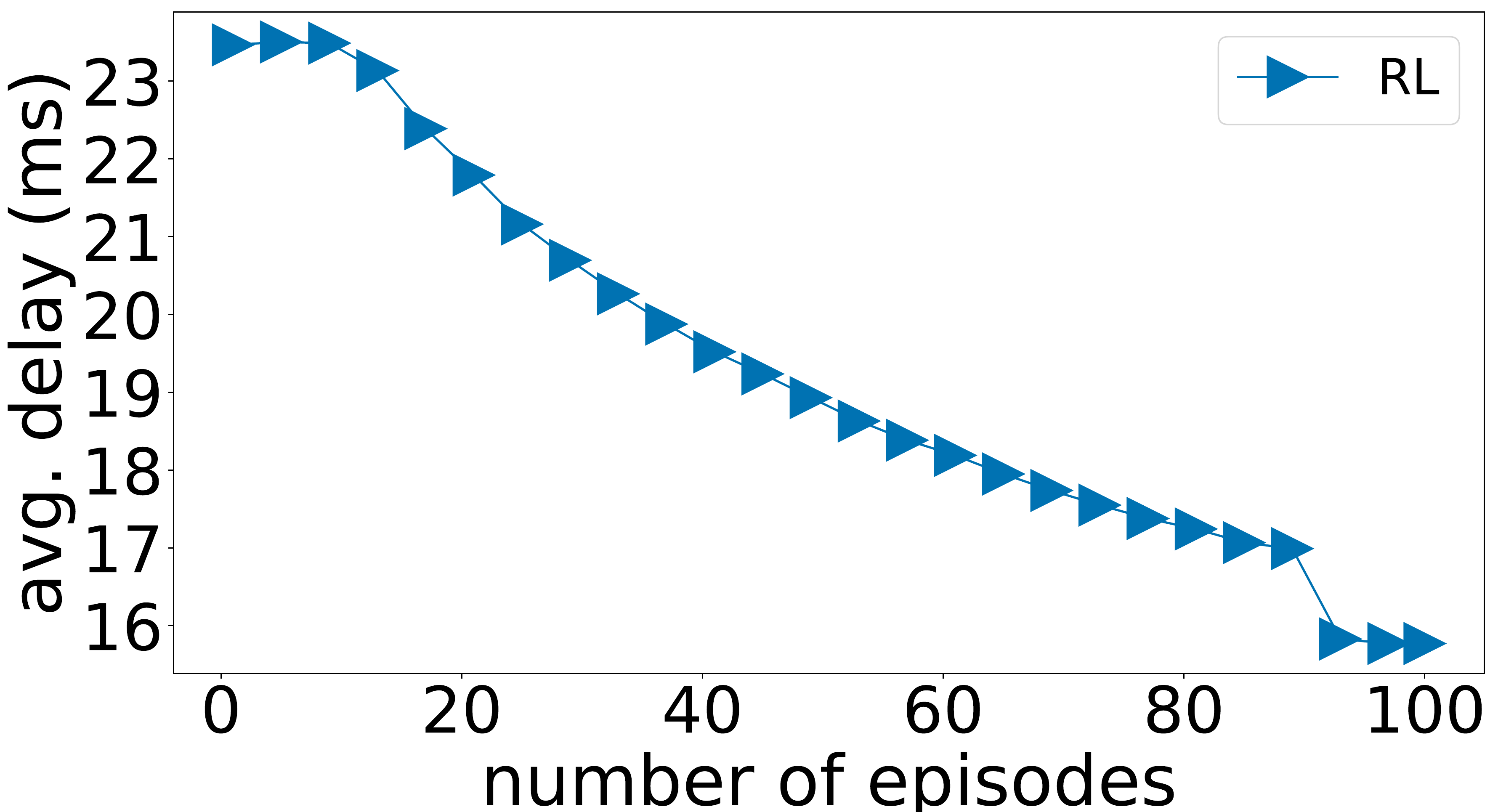}  
\includegraphics[width=0.32\textwidth]{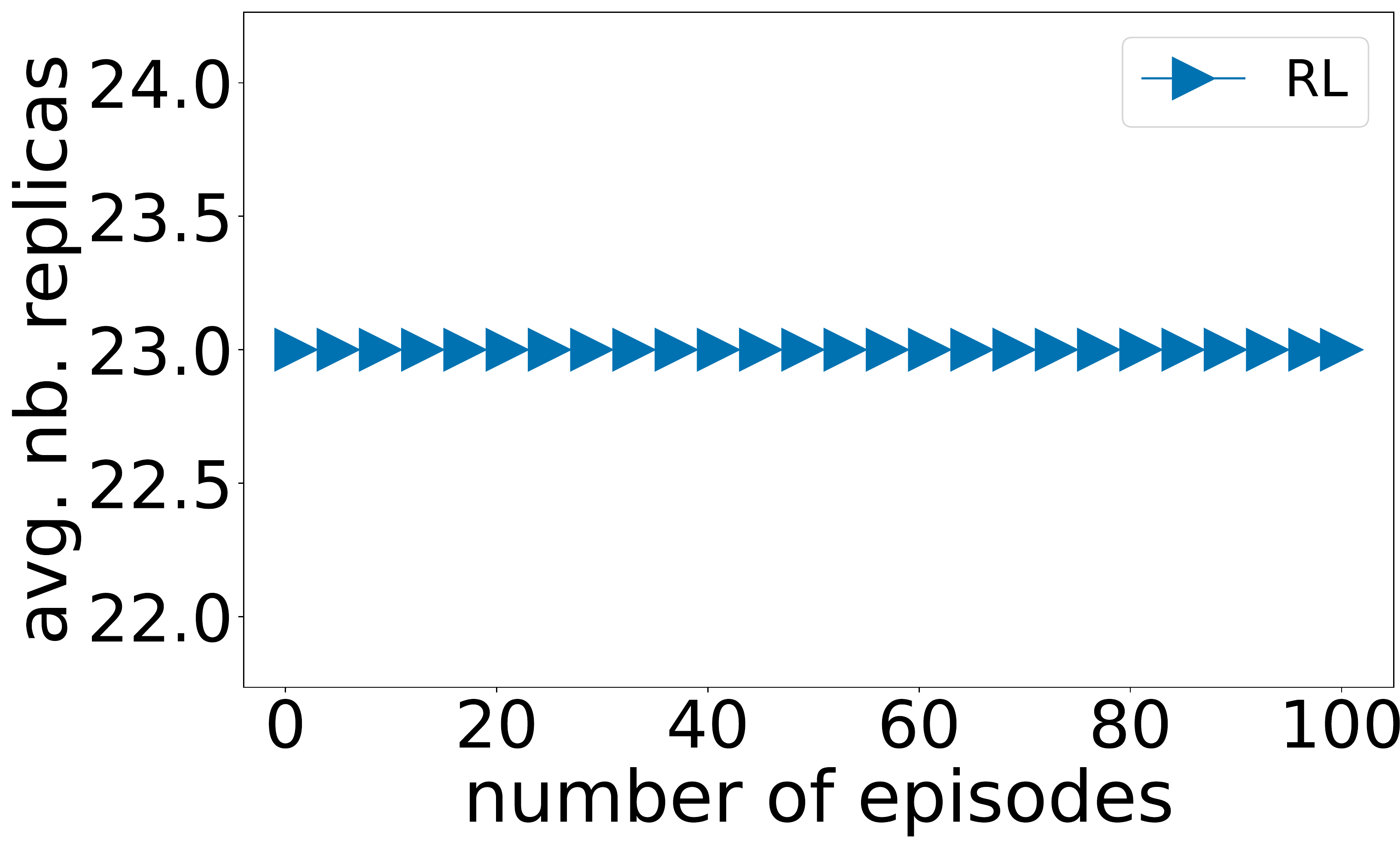}  
   \includegraphics[width=0.32\textwidth]{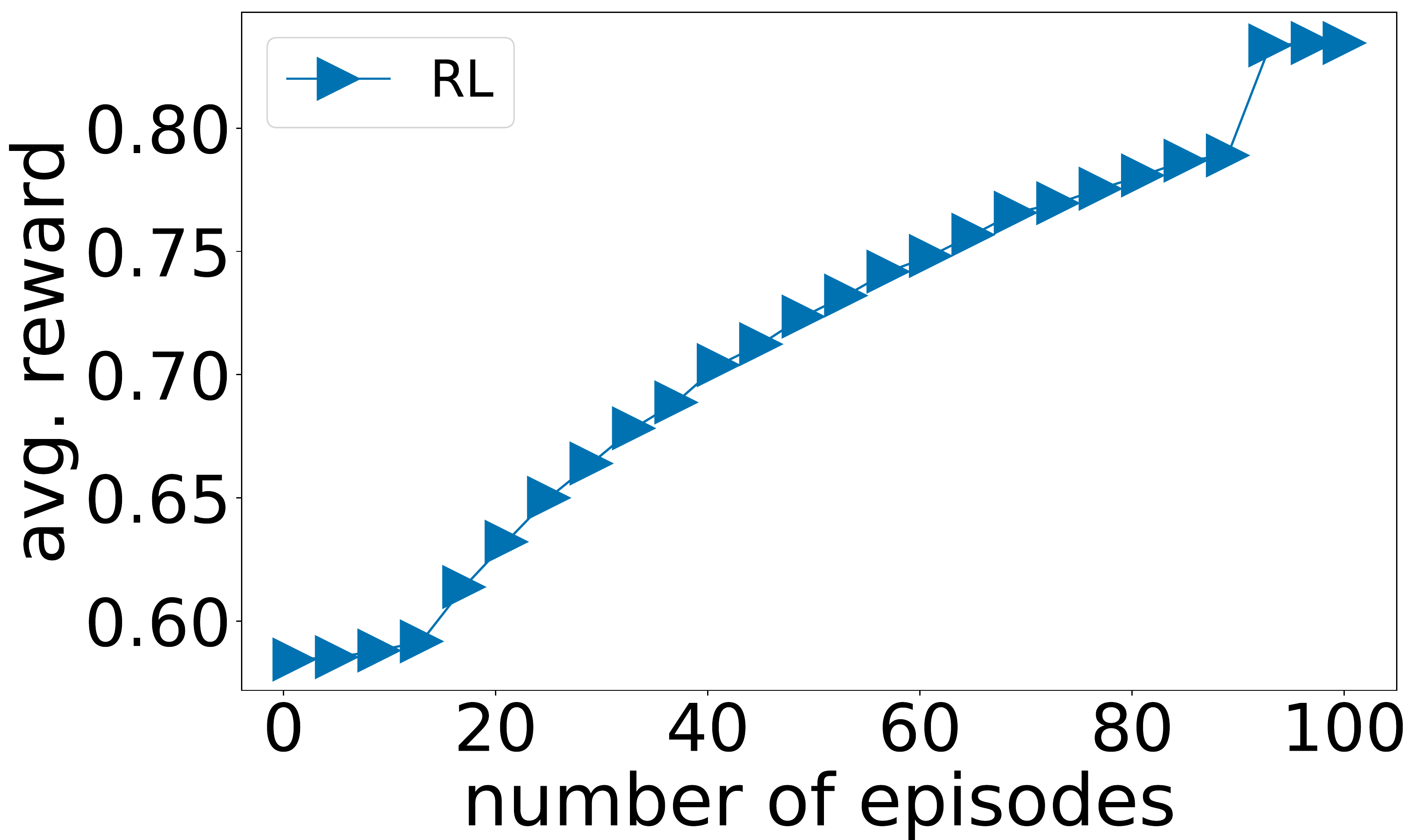}  
\caption{Reinforcement Learning performance per episodes}
\label{RL_ep}
\end{figure*}

\begin{figure*}
\centering
\includegraphics[width=0.32\textwidth]{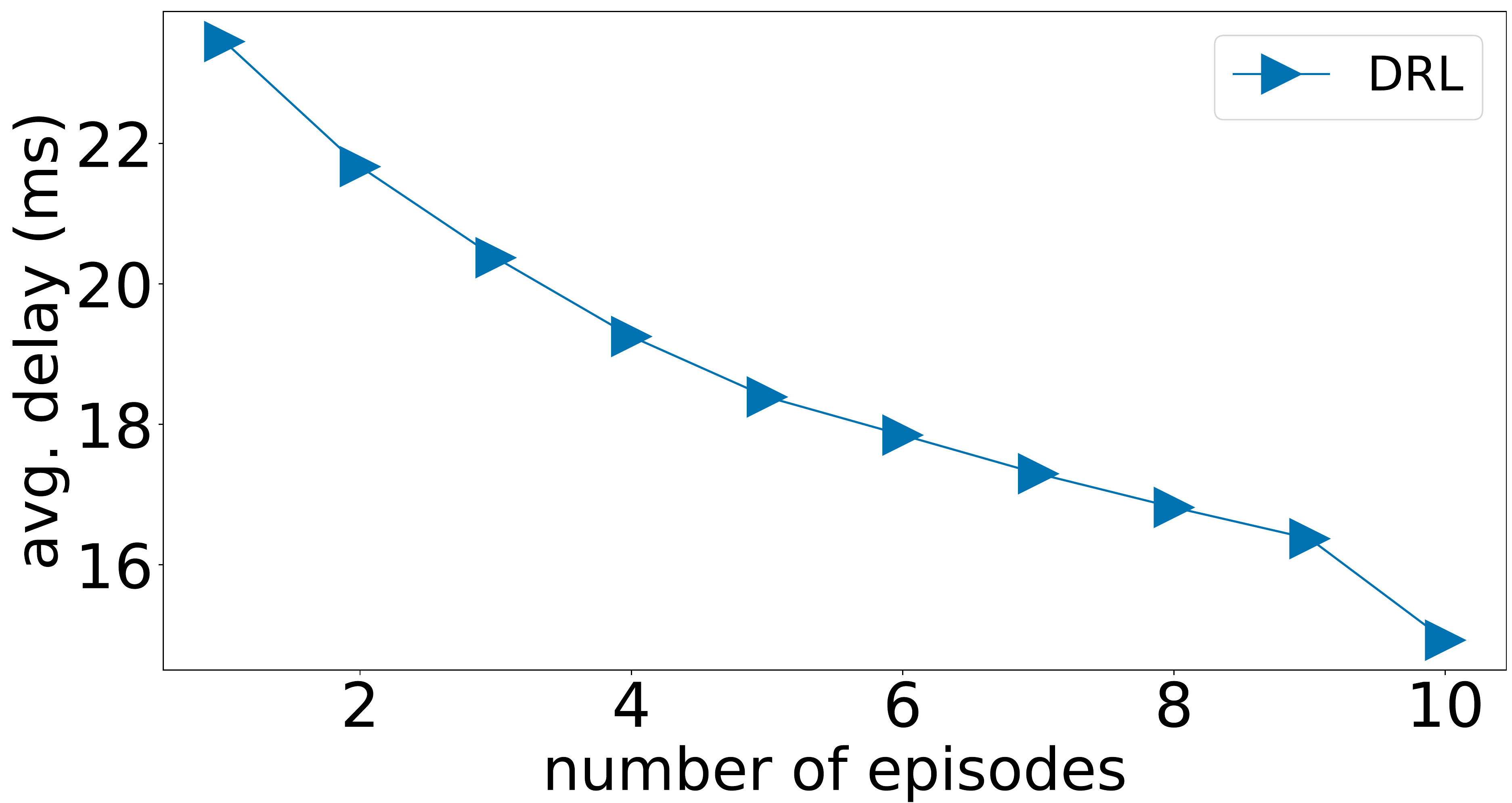}  
\includegraphics[width=0.32\textwidth]{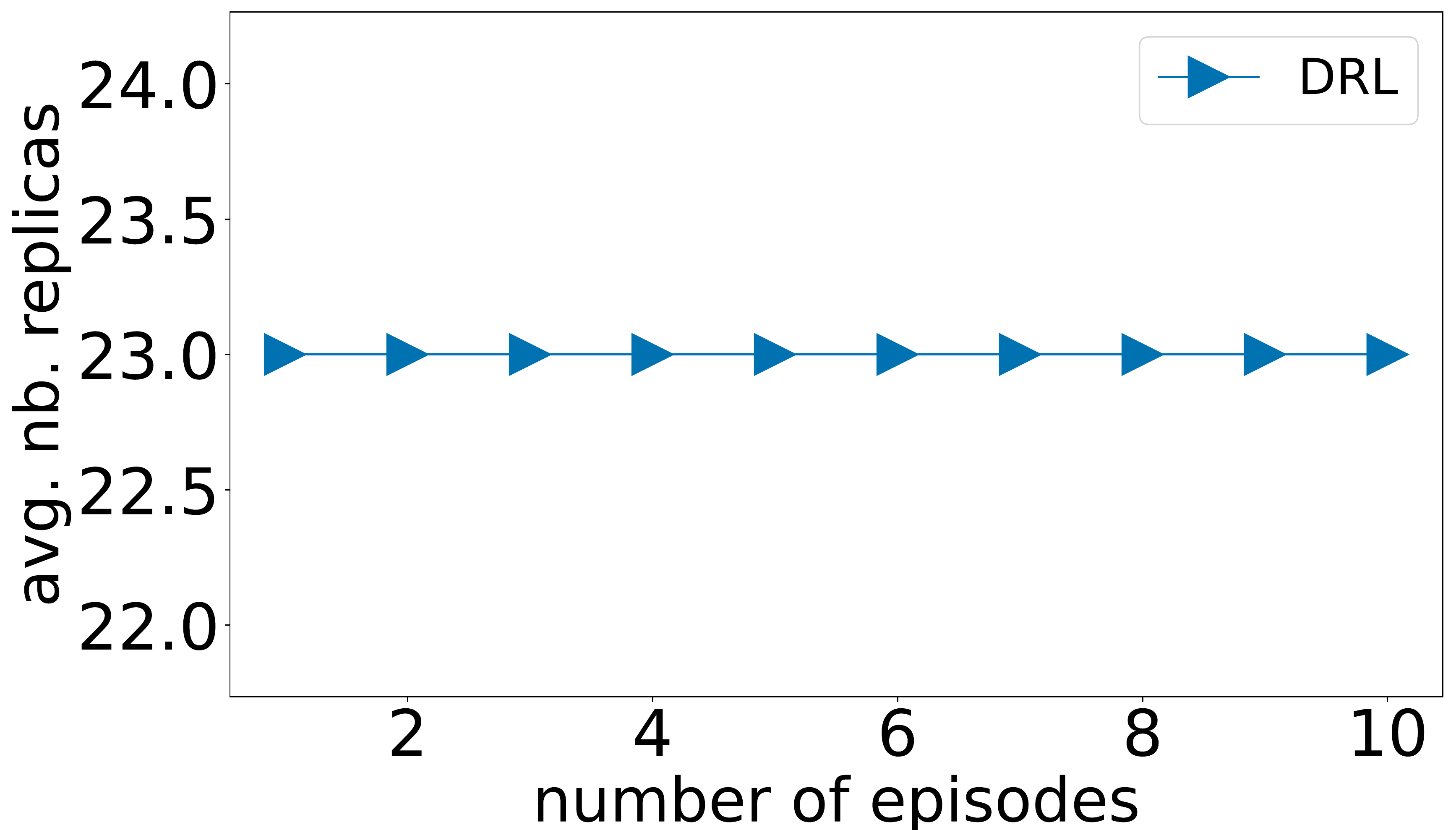}  
   \includegraphics[width=0.32\textwidth]{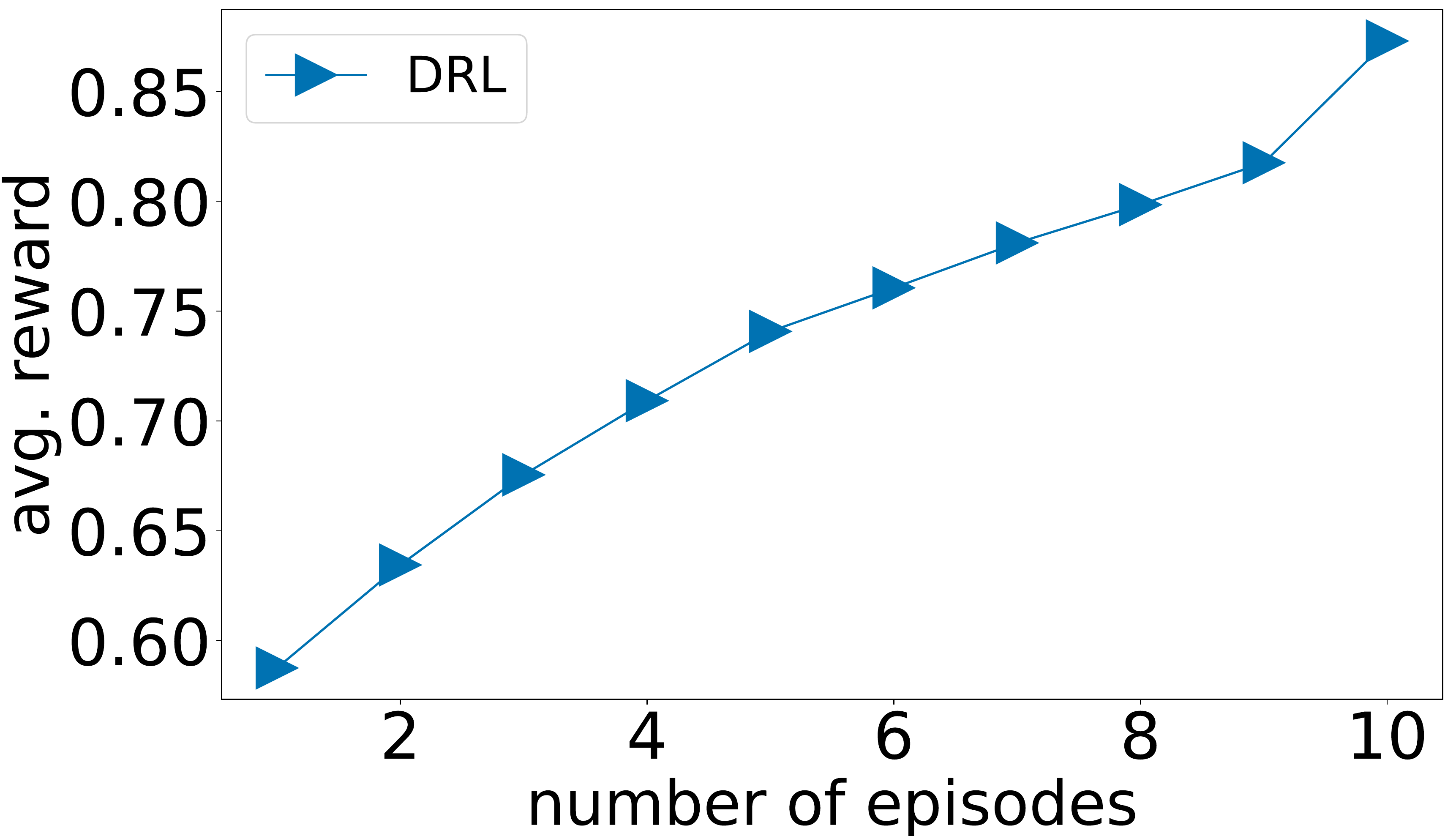}  
\caption{Deep Reinforcement Learning performance per episodes}
\label{DRL_ep}
\end{figure*}

In Figures \ref{RL_ep} and \ref{DRL_ep}, we show the performance results in terms of average delay, average number of replicas, and average rewards, for RL and DRL  algorithms, respectively. The results shows that the average delay starts to decrease linearly after few episodes until it converges at a certain value, which is 90 for RL and 10 for DRL. This proves that the algorithm is learning, from an episode to another, how to efficiently scale functions and place them in the edge nodes. The average delay goes from 23 ms to 16 ms after convergence. The number of replicas remains the same for all episodes, which is expected as the scaling and allocation scheme is satisfying all function requests. The average reward in this case describes the satisfaction rate. i.e. the percentage of satisfied requests in terms of delay constraint (number of delay satisfied requests / total requests). The reward increases with the number of episodes until it converges at the same point of convergence for average delay, to reach a value 90\% and 85\% satisfaction for RL and DRL, respectively.

% DRL analysis
\subsection{Results Discussions}
We now analyze the performance of both RL and DRL scaling methods, by comparing it to  a  monitoring-based heuristics and a delay-aware monitoring approach, in order to verify their efficiency, where the placement is managed by random-fit and first-fit models.  
We adopt the monitoring-based methods for comparison, as they are used by default in Serverless Platforms, such as OpenFaaS. The scaling mechanism in OpenFaaS considers the arrival rate by collecting telemetry about the load. When the load exceeds a predefined threshold, the monitoring  triggers the container orchestrator to create new replicas. When the queue is empty, the monitoring starts to remove replicas, e.g., Kubernetes pods, otherwise the system state remains unchanged. Our delay-aware monitoring approach is an updated version of the same that considers delay constraints of each function, whereby new replicas can be created when the delay constraints are satisfied at least by using one of the existing nodes. The delay aware monitoring algorithm is detailed in \textbf{Algorithm \ref{alg:MBalgorithm}}.

% 

%
%\begin{table}[h]
%	\caption{Q-learning parameters}
%	\label{tab:q_learning_parameters}
%	\centering
%	\begin{tabular}{|c|c|}
%		\hline
%		\textbf{Q-learning Parameter} & \textbf{Value} \\
%		\hline
%		Learning rate ($\alpha$) & 0.01\\
%		\hline
%		Discount factor ($\gamma$) & 0.95 \\
%		\hline
%		Number of episodes & 100 \\
%		\hline
%		Initial $\epsilon$ & 1 \\
%		\hline
%		Decay rate for $\epsilon$-greedy exploration & 0.98 \\
%		\hline
%	\end{tabular}
%	
%\end{table}

\begin{algorithm}
\footnotesize %\small 
\caption{Monitoring-based scaling algorithm}
\label{alg:MBalgorithm}
\begin{algorithmic}[1]
\State \textbf{Input: } event, queue, capacity, load, threshold
\If{event is an arrival }  \If{capacity is available}  \If{for any node $E_n$,  $\xi_k$  is satisfied}  \If{load > threshold}  return 1 
\Else  \;return 0
\EndIf 
\Else  \;return 0
\EndIf
\EndIf
\EndIf
\If{event is a departure }  \If{the queue is empty }   return -1
\Else  \; return 0
\EndIf
\EndIf
\end{algorithmic}
\end{algorithm}

We analyze 4 scaling methods: i) monitoring-based (\emph{MNT}) and ii) delay-aware monitoring-based (\emph{MNT\_constraint}) iii) reinforcement learning (\emph{RL}) and iv) deep reinforcement learning (\emph{DRL}). For i) and ii), the algorithms do not determine function allocations, thus we adopt two simple resource allocation approaches: First-Fit allocation (FFa) and Random-Fit allocation (RFa). In FFa, we allocate  functions in the closest available node. In RFa, we allocate  functions  randomly in any available node with enough available resources. For iii) and iv), the algorithms decide the allocation as well as the scaling.

%\begin{table}[h]
%	\caption{Function parameters}
%	\label{tab:function_parameters}
%	\centering
%	\begin{tabular}{|c|c|c|c|}
%		\hline
%		\textbf{Function} & \textbf{Service Time $\mu_k$} & \textbf{Time Constraint $\xi_k$} & \textbf{Function Req. $b_k$} \\
%		\hline
%		$f_1$ & 5 & 20 & 1 \\
%		\hline
%		$f_2$  & 6 & 23 & 2 \\
%		\hline
%		$f_3$ & 7.828 & 26 & 3 \\
%		\hline
%		$f_4$ & 10.196 & 29 & 4 \\
%		\hline
%		$f_5$ & 13 & 32 & 5 \\
%		\hline
%	\end{tabular}
%	
%\end{table}
% RL analysis

%\begin{figure}[!ht]
%  \centering
%    \includegraphics[width=0.475\textwidth]{figures/All_Plot_as_PDF/Different Arrival Rate/averagedelay_arrivalrate_withoutRR.pdf}
%  \caption{Average service delay per scaling model.}
% \label{fig:avgr-service-delay_small}
% \end{figure}
% 
% \begin{figure}[!ht]
%  \centering
%    \includegraphics[width=0.475\textwidth]{figures/All_Plot_as_PDF/Different Arrival Rate/averagereplicas_arrivalrate_withoutRR.pdf}
%  \caption{Average number of replicas per scaling model.}
% \label{fig:avgr-nb-rep_small}
% \end{figure}
%
% \begin{figure}[!ht]
%  \centering
%    \includegraphics[width=0.475\textwidth]{figures/All_Plot_as_PDF/Different Arrival Rate/averagereward_arrivalrate_withoutRR.pdf}
%  \caption{Average reward per scaling model (small network).}
% \label{fig:avgr-nb-rep_small}
% \end{figure}

\begin{figure*}
\centering
\includegraphics[width=0.32\textwidth]{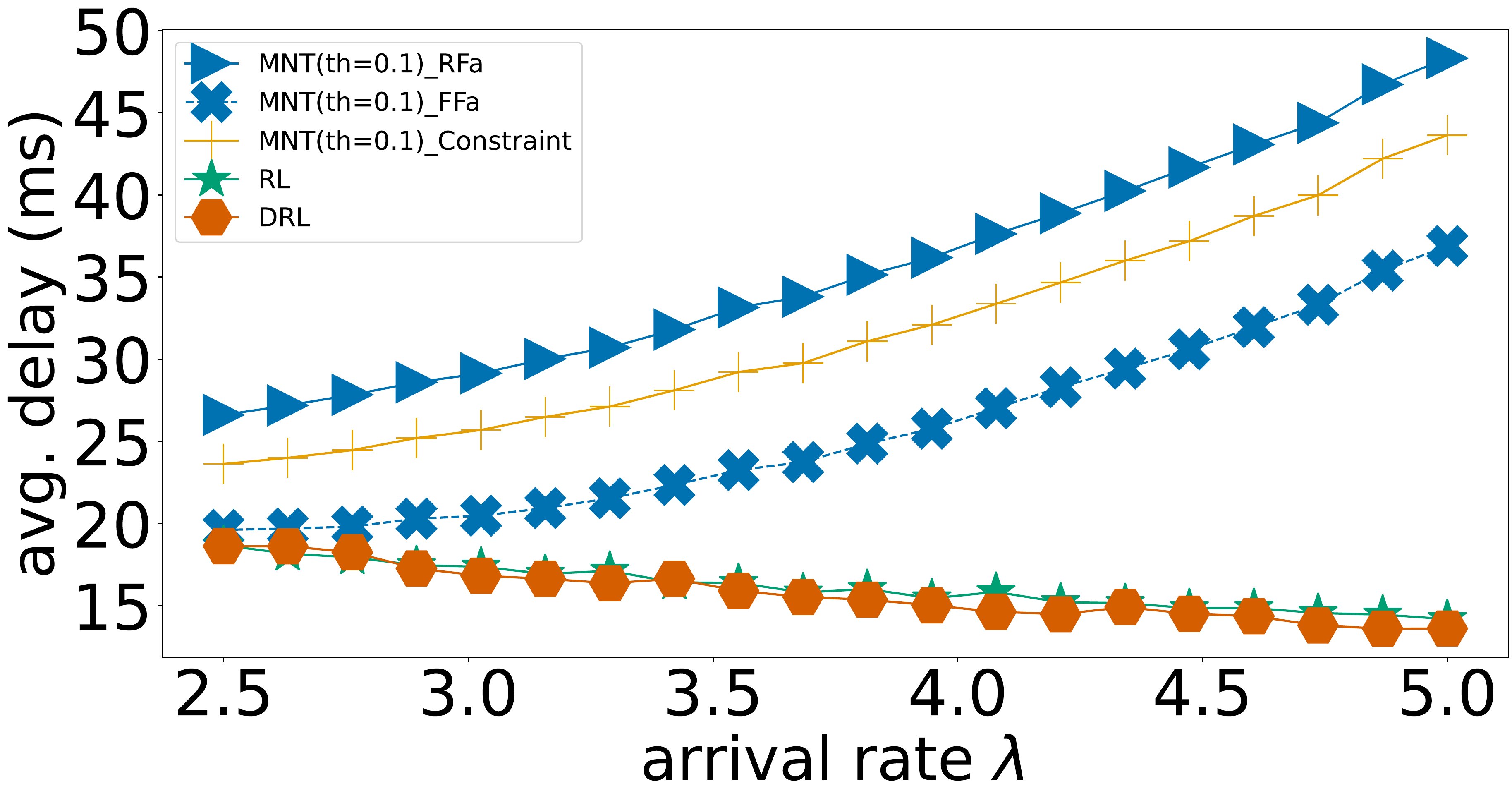}
\includegraphics[width=0.32\textwidth]{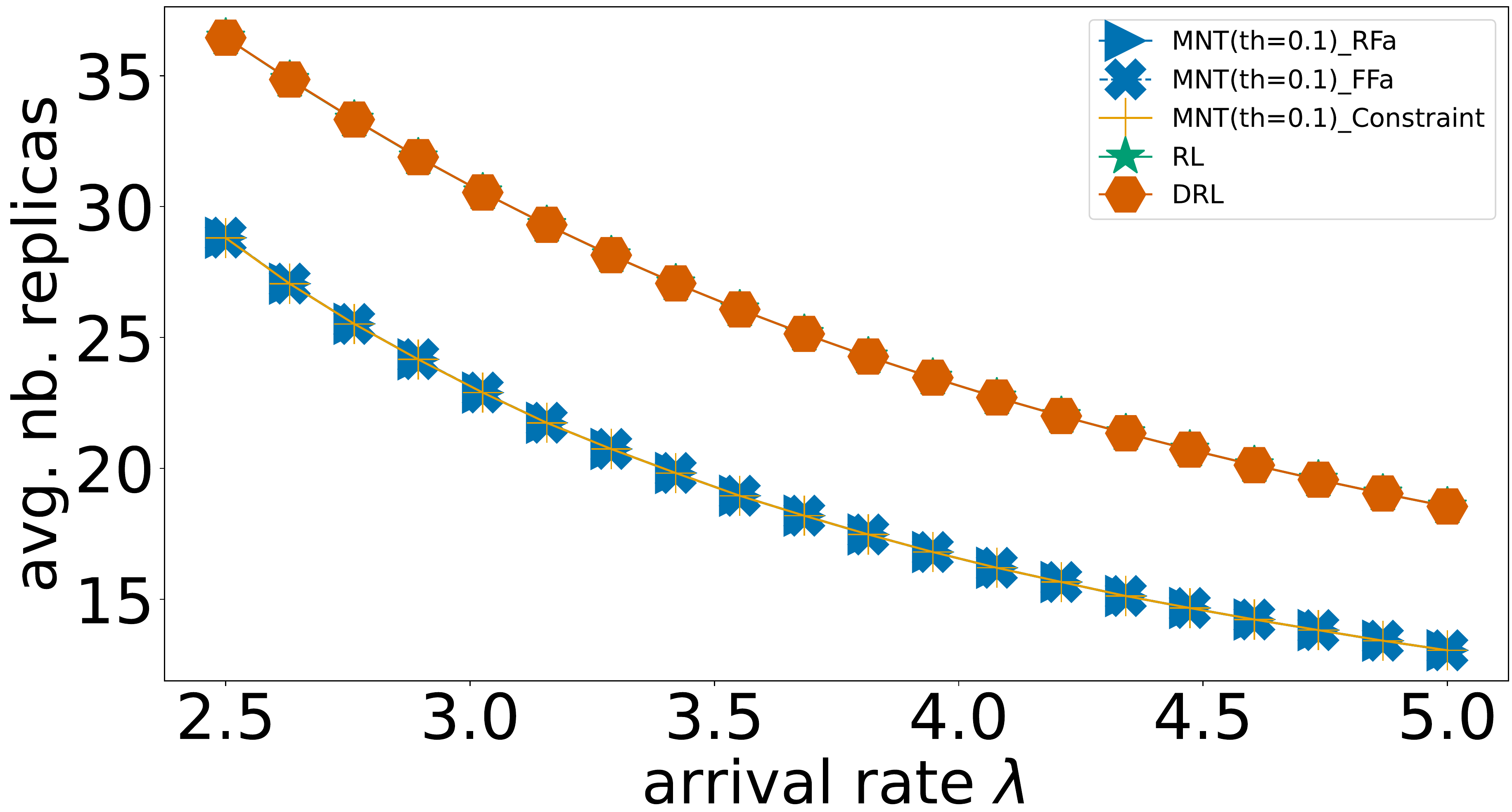}
\includegraphics[width=0.32\textwidth]{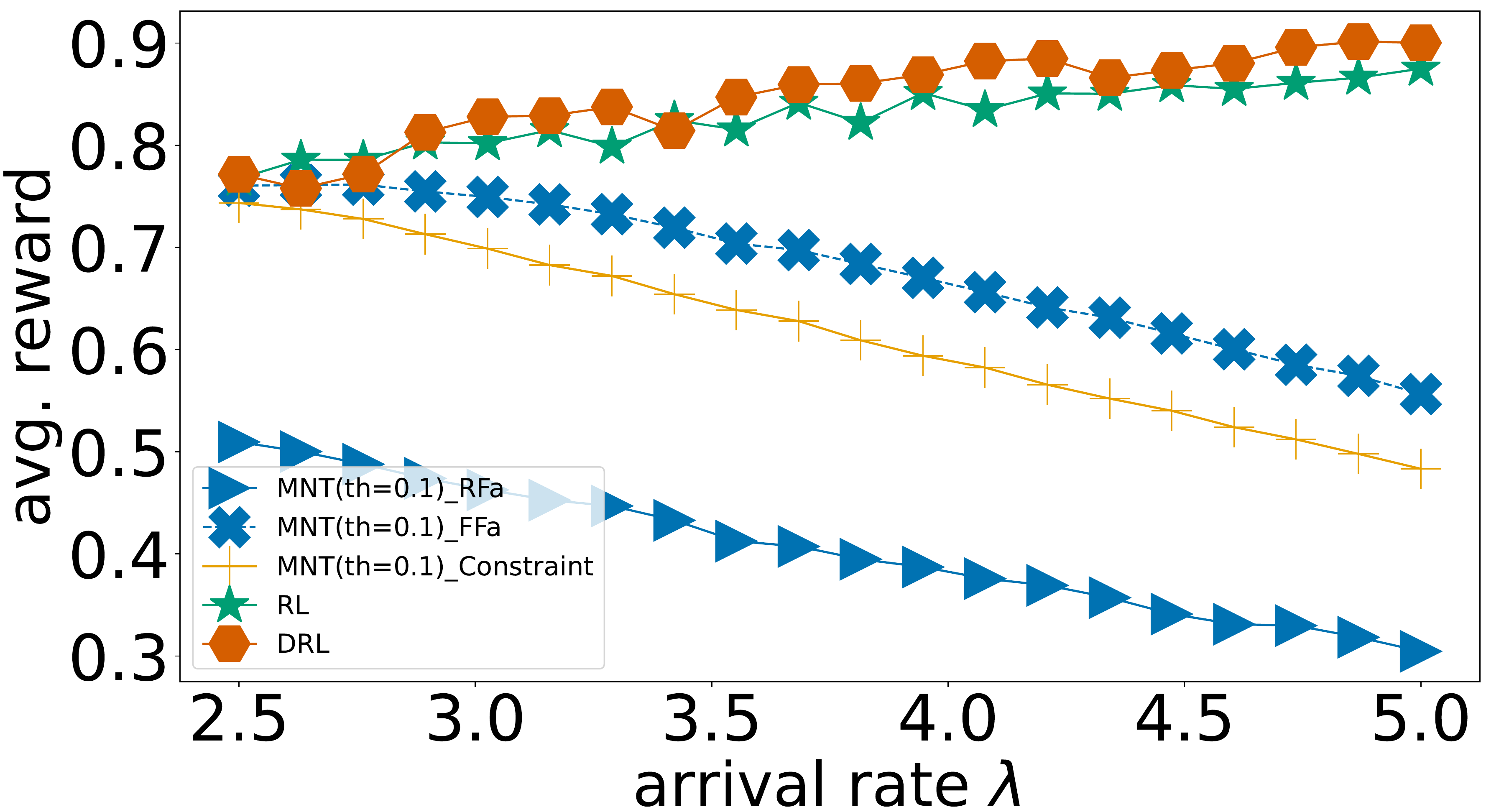}
\caption{Performance results of scaling models per arrival rate.}
\label{fig:perarrival}
\end{figure*}

Figure \ref{fig:perarrival} shows the avg. service delay, the avg. nb. of replicas, and the avg. reward of all functions for arrival rates of functions for each scaling algorithm. The avg. delay results show that RL and DRL outperforms MNT and delay-aware MNT using both first-fit and randon-fit allocation. When the arrival rate increases the load decreases, which decreases the chances of  MNT algorithms to create new replicas. Thus the avg. delay increases with $\lambda$. Meanwhile for RL and DRL, the reward function allows the learning approach to allocate functions in nodes that satisfy delay constraints and decrease the avg.  delay, which explains the decrease of the delay with the decrease of the load (increase of $\lambda$). The delay-aware MNT performs worse than MNT in terms of delay as it may take conservative decision of scaling when the delay constraint is not satisfied. In terms of nb. of replicas, RL and DRL are less conservative than MNT, in order to load balance servers and decrease delays. In terms of reward/ delay satisfaction rate, RL and DRL improves the satisfaction by 50\% comparing to MNT when $\lambda=5$. MNT algorithm performs worst when $\lambda$ increases, as explained for the delay, while RL and DRL performs better as the load decreases. Results given by both RL and DRL are very close, which proves that basic RL is learning efficiently without the need for complex NN models. 
%%%%%%%%%%%%%%%%%%%%%%%%%%%%%%%%%%%%%%%%%%%%%%

\begin{figure*}
\centering
\includegraphics[width=0.32\textwidth]{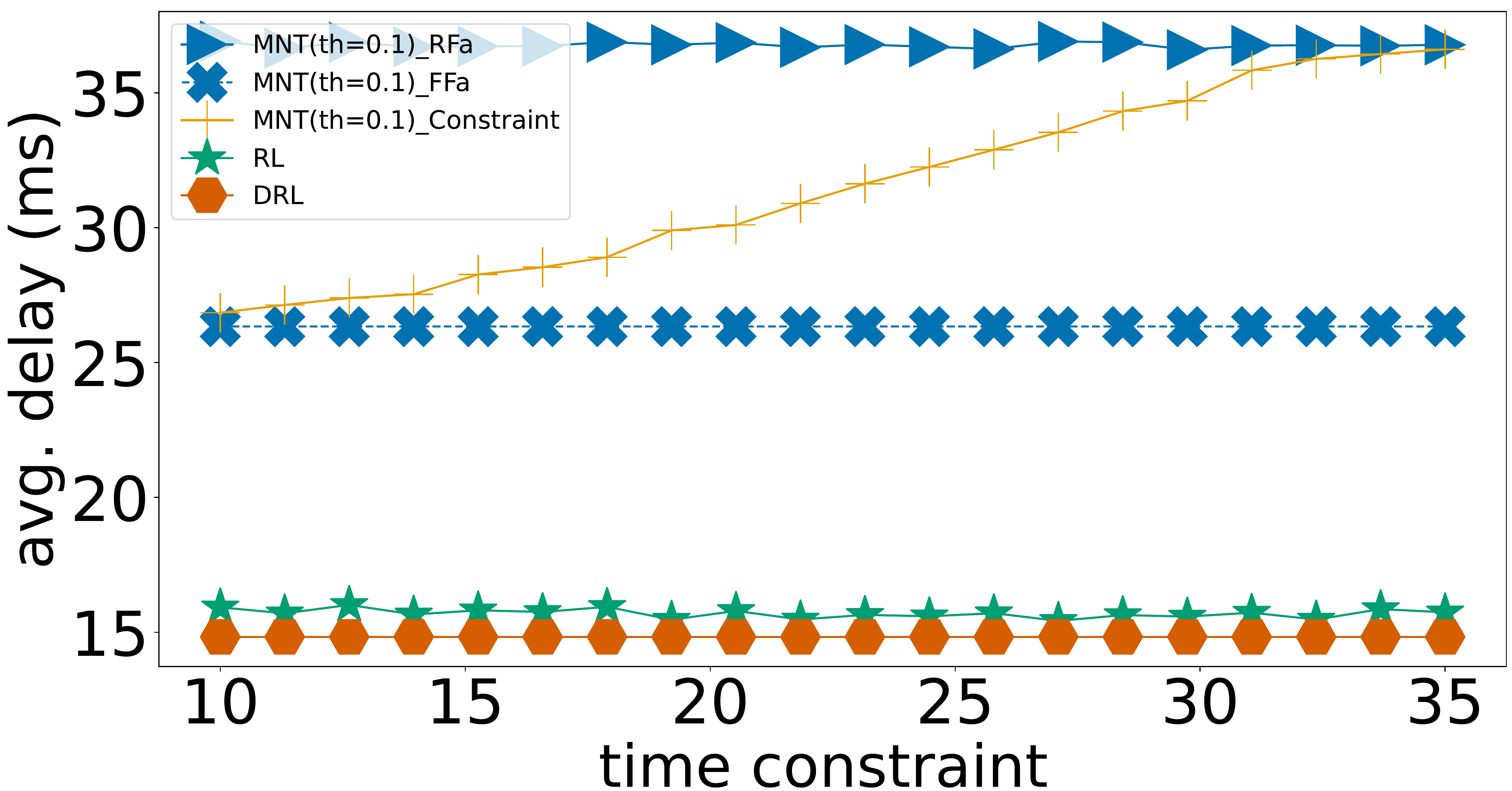}
\includegraphics[width=0.32\textwidth]{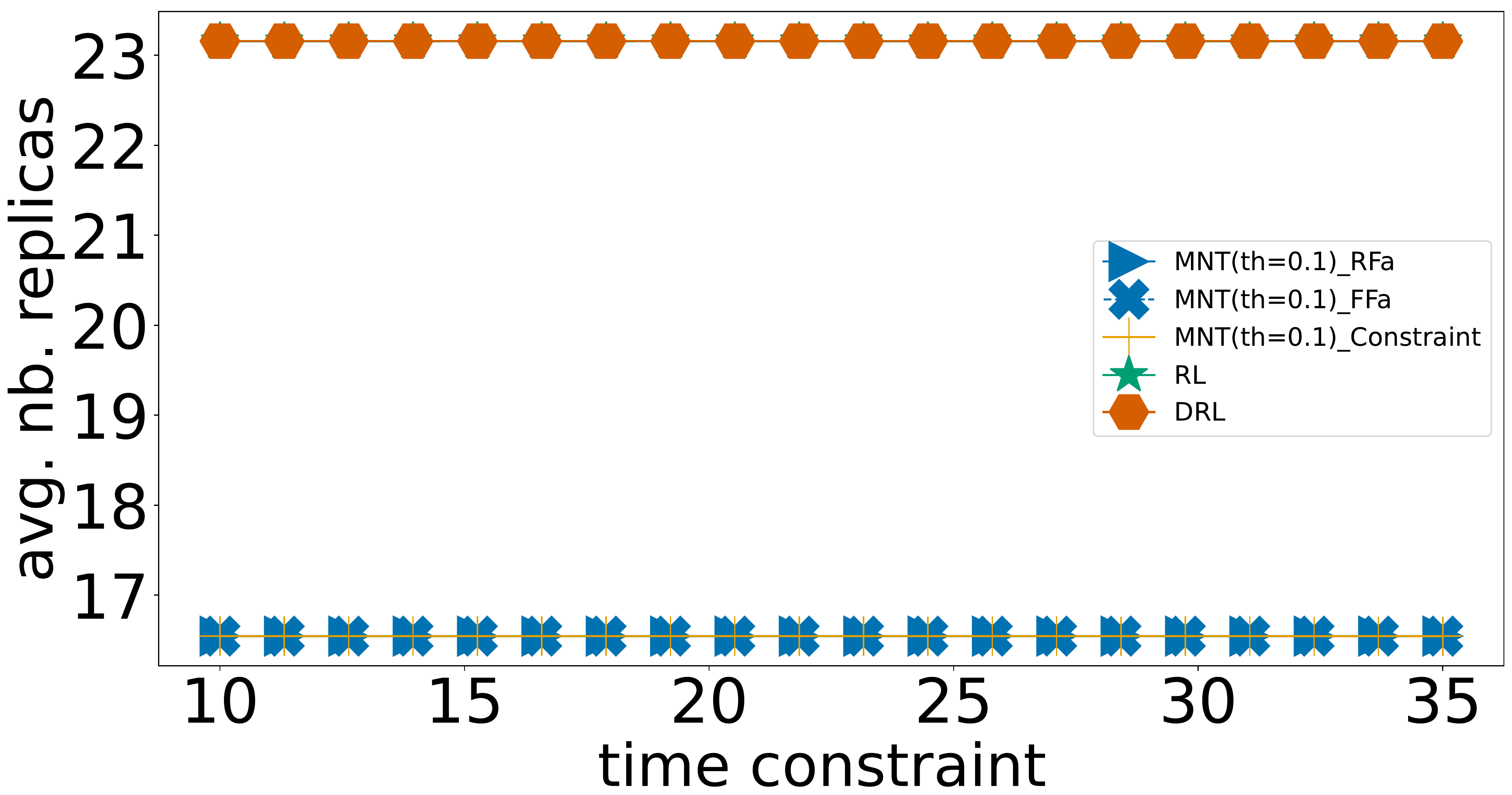}
\includegraphics[width=0.32\textwidth]{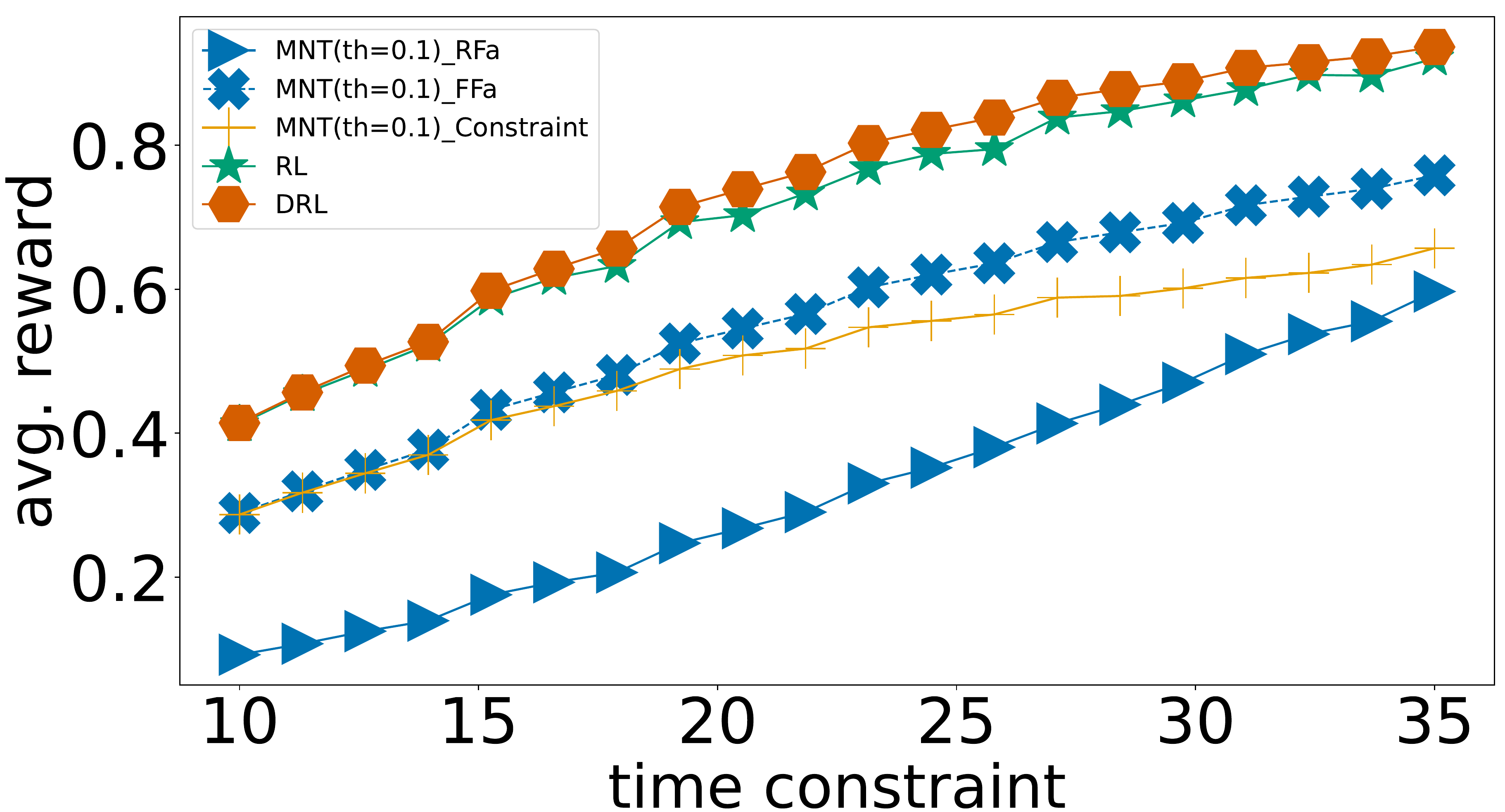}
\caption{Performance results of scaling models per time constraint.}
\label{fig:perdelay}
\end{figure*}

Figure \ref{fig:perdelay} shows the avg. service delay, the avg. nb. of replicas, and the avg. reward of all functions for time constraints  of functions for each scaling algorithm. RL and DRL  outperforms MNT approaches in terms of avg. delay for all values of delay constraints. The results obtained remains the same for all the constraint values, which means that the allocation decision is not affected by constraints, and by incorporating a delay term in the reward function, RL and DRL give the lowest possible avg. delay, while for MNT, the constraint is not considered. For delay-aware MNT, the algorithm becomes very conservative in terms of scaling when the delay constraint increases. In terms of nb. of replicas, RL and DRL use a higher nb. of replicas and for all algorithms, this value is not affected by the delay constraint. For the reward, RL and DRL give higher satisfaction rate for all values, and the rate increase with the decreasing of delay constraint.
\vspace{-0.2cm}
 \section{Conclusion}\label{sec:conclusion}
In this paper, we addressed  the problem of efficient scaling and resource  allocation of serverless functions in edge networks, considering delay-sensitive applications by applying reinforcement learning (RL) approach.
 We compared RL and Deep RL algorithms with  monitoring-based heuristics, used for practical solutions. The simulation results shows that RL algorithm outperforms the default, monitoring-based algorithms, regarding the total delay of function requests, and achieves an improvement in terms of delay performance by up to 50\%.  Results showed that  RL is as good as DRL, which allows us to use the basic RL as a fast and efficient solution. 
 As a future work, we will evaluate our solution for more complex scenarios, and consider the reliability of edge nodes.

% 
%In this paper, we addressed the optimality of serverless scaling in edge computing network and proposed to use  Semi-Markov Decision Process-based (SMDP) model for the scaling problem of serverless functions as a decision making problem, with actions of scaling the functions up or
%down. The theoretical results were compared with practical, monitoring-based algorithms based on current approaches.The results confirmed that SMDP gave best results in terms of queuing delay, and outperformed
%monitoring-based approaches. The monitoring-based
%approach however achieved performance comparable to the optimal SMDP solution in terms of
%delay when the scaling activation threshold was set to comparably lower values.% \hl{we need one more sentence on resutls?! and before going to future work, explain the shortcoming of thsi approach. what are the badn things we found and how to fix them. then finish with future work.} 
% In our future work, we will study the joint scaling and resource allocation problem
%considering various system parameters. SMDP can be used to analyse the optimal
%policy for a specific setting. Monitoring based is what is used in real
%implementations now, which can be more optimized using smarter algorithms such
%as SMDP. The issue of SMDP is its exponential expansion, so a heuristic based on
%SMDP can be a very good approach for
%future.% but it is not proposed in this paper 

%15:26 smdp and monitoring based approach can be combined.
% SMDP already assumes that some information is known such as the arrival rate and service rate of each function.
%

\vspace{-0.2cm}

%This work was partially supported by the DFG Project Nr.
%JU2757/12-1, and by the Federal Ministry of Education and
%Research of Germany, joint project 6GRIC, 16KISK031.
%
%
% This work is partially funded by European Commission
%under the H2020-952644 contract for project FISHY: A coordinated framework for cyber resilient supply chain systems
%over complex ICT infrastructures.

\section*{Acknowledgment}
%This work was partially supported by EU HORIZON research and innovation program, project ICOS, Grant Nr. 101070177,  and by  European Commission for project FISHY under the H2020, Grant Nr.952644.
This work was partially supported by EU HORIZON research and
innovation program, project ICOS, Grant Nr. 101070177,  and by  European Commission for project FISHY under the H2020, Grant Nr.952644.

\vspace{-0.4cm}
\bibliographystyle{IEEEtran}
\bibliography{mybib}

\end{document}